\documentstyle[twocolumn,aps]{revtex}

\newcommand{\beq}{\begin{equation}}
\newcommand{\eeq}{\end{equation}}
\newcommand{\beqa}{\begin{eqnarray}}
\newcommand{\eeqa}{\end{eqnarray}}
\newcommand{\beqar}{\begin{eqnarray*}}
\newcommand{\eeqar}{\end{eqnarray*}}

\begin{document}
\title{{\bf {\Large A classical analog to topological non-local quantum
interference effect}}}
\author{Yakir Aharonov$^{a,b}$, Sandu Popescu $^{c,d}$, Benni Reznik$^{a}$
and
Ady
Stern$^e$ {\ }}
\address{(a) {\em {\small School of Physics and Astronomy, Tel Aviv
University, Tel
Aviv 69978, Israel}}\\
(b) {\em {\small Department of Physics, University of South Carolina,
Columbia, SC 29208}}\\
(c) {\em {\small H.H. Wills Physics Laboratory, University of Bristol,
Tyndall Avenue, Bristol BS8 1TL, UK}} \\
(d) {\em {\small Hewlett-Packard Laboratories, Stoke Gifford, Bristol BS12
6QZ, UK}}\\
(e) {\em {\small Department of Physics, Weizmann Institute of Science,
Rehovot 76100, Israel}}}
\maketitle

\begin{abstract}
{The two main features of the Aharonov-Bohm effect are the topological
dependence of accumulated phase on the winding number around the magnetic
fluxon, and non-locality -- local observations at any intermediate point
along the trajectories are not affected by the fluxon. The latter property
is usually regarded as exclusive to quantum mechanics. Here we show that
both the topological and non-local features of the Aharonov-Bohm effect can
be manifested in a classical model that incorporates random noise. The model
also suggests new types of multi-particle topological non-local effects
which have no quantum analog. }
\end{abstract}




In the Aharonov-Bohm (AB) effect\cite{AB}, a phase
\begin{equation}
\phi_{AB} =\hbar q\oint \vec A \cdot \vec dl,
\end{equation}
is accumulated by a charge $q$ upon circulating a solenoid enclosing a
magnetic flux. This phase (mod $2\pi$) can be observed in an interference
experiment in which the wavefunction of the charge is split into two
wavepackets which encircle the solenoid and then interfere when meet
together. The phase is {\it topological} because it is determined by the
number of windings the charge carries out around the solenoid, and is
independent of the details of the trajectory. The phase is also {\it %
non-local}: while the magnetic flux in the solenoid clearly affects the
resulting interference pattern, it has no local observable consequences
along any point on the trajectory. There is no experiment that one can
perform anywhere along the trajectory which can tell whether or not there is
a magnetic flux inside the solenoid. (In particular, there is no force
acting on the charge due to the solenoid.)

Various topological analogs of the AB effect have been suggested utilizing
light in an optical medium\cite{milloni}, super-fluids\cite{super} and
particles in a gravitational background\cite{gravity}. However unlike the AB
effect, in these analogs one can observe how the global topological effect
builds up locally. Hence these models do not reproduce the non-local aspect
of the AB effect. (The gravitation analog is an exceptional case. However it
requires a non-trivial space-time structure, which is locally flat but
globally not equivalent to a Minkowski space-time.)

The above analogs suggest that quantum systems differ fundamentally from
classical systems as far as non-locality is concerned. In this letter
however we show that a classical non-local effect may be constructed without
employing a non-trivial space-time structure. The new ingredient which
allows for this is the inclusion of a random bath of particles, which
``masks" local effects, but does not ``screen'' the net topological effect.

The classical non-local effect we are constructing now is a classical analog
of the Aharonov-Casher effect\cite{AC}. To begin with, consider a particle
that is described by the canonical coordinates ${\cal \vec R}$ and ${\cal
\vec P}$%
, and is carrying a magnetic moment $\vec \mu$. Let this particle,
(henceforth referred to as the fluxon) interact with the electric field $%
\vec E$ generated by a homogeneously charged straight wire positioned along
the $\hat z$-axis. The non-relativistic Hamiltonian describing this system
is the one employed in the study of the Aharonov-Casher effect
\begin{equation}
H_{AC} = {\frac{(\vec {{\cal P}} + \frac{1}{c}\vec\mu\times \vec E)^2}{2m}}.
\end{equation}

For simplicity, in the following we confine the motion of the fluxon to a
two dimensional plane orthogonal to the wire and effectively reduce the
system to be planar. We denote by $\vec R$, $\vec P$ the two dimensional
position and momentum, respectively. We assume that a time dependent scalar
potential is applied to the fluxon to make it move along a desired
trajectory in the plane, and that this potential does not interact with the
magnetic moment. For brevity we do write this potential below, as it does
not affect our considerations. Using polar coordinates, the electric
field can be written as $\vec E(R) = 2\lambda|\vec \nabla_R\theta| \hat R$,
where $\lambda$ is the linear charge density and $\vec\nabla_R$ is the
gradient
with respect to $\vec R$ and $\hat R$ is the unit vector along $\vec R$. If
$%
\vec \mu$ is aligned in the $\hat z$ direction, the Hamiltonian becomes
two-dimensional\cite{point-Q}
\begin{equation}
H_{AC} = {\frac{(\vec P + {\frac{2}{c}}\mu \lambda \vec \nabla_R\theta
)^2}{%
2M}}.  \label{hac}
\end{equation}

Classically, the forces on the particle vanish. However the magnetic field
experienced by the fluxon in its rest frame, $\vec B = \frac{\vec v \times
\vec E}{c}$ is non-vanishing. Therefore one expects a non-zero torque, $\vec
\mu \times \vec B$, and consequently an ``internal'' precession of the
magnetic moment. This latter effect is present both in the quantum and the
classical cases \cite{peshkin-lipkin,aharonov-reznik}. The precession is not
described by the above Hamiltonian because $\mu$ was taken as fixed vector,
rather than a degree of freedom. To incorporate the internal precession we
next introduce an internal angular momentum variable $\vec L=\hat z L$, with
a conjugate internal angular coordinate $\varphi$. We further assume that $%
\mu\propto L$ (Indeed the magnetic moment of a neutron, for example, $\mu_N
= -3.7{\frac{e}{2mc}}{\bf s}$, is proportional to its spin $s$.) Replacing
$%
\mu$ in eq. (\ref{hac}) by $L$ we have

\begin{equation}
H = {\frac{(\vec P + \xi L \vec\nabla_R\theta )^2}{2m}}  \label{hl}
\end{equation}
where $\xi$ is the resulting net charge/magnetic-moment coupling constant.

While $L$ (and the magnitude of $\mu$) are constants of motion, the internal
angle $\varphi$, conjugate to $L$, varies in time and satisfies the equation
of motion
\begin{equation}
{\frac{d\varphi}{dt}} = \xi{\frac{d\vec R}{dt}} \cdot\vec\nabla_{R} \theta =
\xi%
{\frac{d\theta}{dt}}.  \label{eom}
\end{equation}
Consequently $\varphi$ is entirely determined by the polar angle $\theta$ of
the fluxon relative to the $x$-axis emanating from the position of the
charged line\cite{note1}. Once we fix the initial value $\varphi(\theta_0)$
for a given $\theta_0$, the internal coordinate $\varphi$ at a later time is
given by
\begin{equation}
\varphi(\theta) = \xi(\theta -\theta_0) + \varphi(\theta_0).
\end{equation}
Hence as the fluxon moves along a closed loop enclosing the charge, and $%
\theta$ changes by $2\pi$, the internal angle changes by $2\pi\xi$. Below we
refer to the case where $\xi$ is interger as ``trivial'', as it leads to an
angle winding of a multiple of $2\pi$, and to the case of $\xi$ non-integer
as ``non-trivial''. This situation is similar to the experimentally observed
Aharonov-Casher effect in a Josephson junction array \cite{Elion}.

Up to now we have constructed a model which exhibits a classical analog of a
topological effect. However since we can locally observe how the internal
angle changes at intermediate points of the trajectory, this model does not
capture the non-local feature of an AB-like effect.

As we now turn to show, an effectively non-local behavior emerges if we add
to the above system two new ingredients. Firstly, we employ two fluxons.
Secondly, we consider the interaction of these fluxons with a non-trivial
charged particle situated at the origin (i.e. a particle with non-integer $%
\xi$), and a bath of randomly positioned, moving, charged particles, all
leading to trivial angle windings of the fluxon (i.e. particles with integer
$\xi$). As each fluxon encircles the origin, the particles of the bath
randomize its angle. These particles do not, however, randomize the angle
{\em difference} between the two fluxons when they coincide in position.

Let us denote the coordinates of the two fluxons by $\vec R_k, \vec P_k$,
and internal coordinates by $\varphi_k$ and $L_k$. The coupling constants
with the particles of the bath are taken to be be ``trivial'', i.e., $%
\xi_{1i}=\xi_{2i}=1$. We denote the coordinates of the bath particles by $%
\vec r_i=x_i\hat x + y_i\hat y$ and their momenta by $\vec p_i$ (with $%
i=1...N$). The Hamiltonian of the system becomes,
\begin{eqnarray}
H &=& \sum_{k=1}^2{{\frac{(\vec P_k + L\vec \nabla_{R_k} [\xi \theta_k
+\sum_i \theta_{ki}] )^2 }{2M}}}  \nonumber \\
&+&\sum_{i=1}^N {\frac{ (\vec p_i + L\sum_{k=1}^2\vec
\nabla_{r_i}\theta_{ik} )^2
}{2m_i}}.
\end{eqnarray}
The first term above represents two fluxons, which interact with the charged
particle at the origin and with the bath. The second term represents this
``bath''. Notice that the kinetic term for the charged particles of the bath
includes a vector potential, too\cite{chargehamiltonian}. The presence of
the bath exerts additional vector potential terms
\begin{equation}
\vec A_{ki}= L\vec \nabla_{R_k}\theta_{ki}
\end{equation}
where $\theta_{ik} =\arctan{\frac{y_i-Y_k}{x_i-X_k}}$, is the angle between
$%
\vec r_i -\vec R_k $ and the $x$ axis.

As we have seen above (Eq. (\ref{eom})) the internal angle changes according
to the relative angle between the fluxon and the charged particle. Indeed,
the equation of motion for the internal angle is
\begin{eqnarray}
{\frac{d \varphi_k}{dt}} &=& \xi {\frac{d\vec R_k}{dt}}\cdot \vec\nabla%
_{R_k}\theta_{k}+ \sum_i\biggl({\frac{d\vec R_k}{dt}}\cdot \vec \nabla_{R_k}
+ {\frac{d \vec r_i}{dt}} \cdot \vec \nabla_{r_i}\biggr) \theta_{ki}
\nonumber \\
&=& \xi {\frac{d\theta_{k}}{dt}} + \sum_i {\frac{d\theta_{ki}}{dt}}.
\end{eqnarray}

Clearly, for a sufficiently large number of randomly distributed particles,
the effect of the bath on the internal angle becomes chaotic. The time
dependence of the $\varphi(t)$ becomes unpredictable.

Consider now however the following experiment. We start with the two fluxons
situated at the same point. Then one of the fluxons stays fixed while the
other moves in a path around the non-trivial charge and returns to its
initial point. As noted above, the internal angles of each fluxon change
randomly. But consider the the {\it relative} internal angle between the
fluxons,
\begin{eqnarray}
\gamma(t) & \equiv &\varphi_2(t)-\varphi_1(t) \\
&=& \xi(\theta_{1}(t)-\theta_{2}(t)) + \sum_{i\in bath} (\theta_{i1}(t)
-\theta_{i2}(t)) + constant.  \nonumber
\end{eqnarray}

We first note that when the two fluxons are located at precisely the same
point, $\vec R_1= \vec R_2$ we have
\begin{equation}
\theta_{i1}(t) = \theta_{i2}(t).
\end{equation}
Therefore the random changes induced by the bath in the internal angles of
the fluxons are {\em identical}, and as long as the fluxons coincide
\begin{equation}
\gamma(t)=constant .
\end{equation}

Once the fluxons move apart, the random time dependence of $\varphi_1(t)$
differs from that of $\varphi_2(t)$, and $\gamma(t)$, the relative internal
angle, becomes random.

Finally however, when the moving fluxon returns to its original point, after
$n$ windings around the origin, and the two fluxons coincide again,
\begin{equation}
\gamma_{final} -\gamma_{initial}= 2\pi n\xi + 2\pi N .
\end{equation}
The first term is the shift caused by the charge at the origin. The second
term is due to the bath and $N$ is an integer random
number which counts the number of windings of bath particles.
The particles of the bath can
wind around only one of the fluxons while the fluxons are apart.
However when the fluxons coincide, the final relative winding number $N$
is a random integer. More importantly, $(\gamma_{final}-%
\gamma_{initial}) mod 2\pi$ is unaffected by the bath particles, in sharp
contrast to the values of $\varphi_1,\varphi_2,\gamma$ along the trajectory.
Thus, upon closing a loop, the random effects due to the bath particles
cancel and $(\gamma_{final}-\gamma_{initial}) mod 2\pi$ depends only on the
non-trivial charge. In other words, although during the experiment the
internal angles change randomly, upon closing a loop and measuring the
change of the internal angle of one fluxon with the respect to the other
fluxon (which acts as a reference system) we are able to recover information
about the non-trivial charge. The effect is {\it topological}, because it
depends only on the winding numbers and not on the details of the loop.
Furthermore, and most important, the effect is {\it non-local} because no
useful information can be extracted on a local basis (i.e. by monitoring the
changes only on parts of the loop); only the closed loop yields information.

More generally, we can allow both fluxons to move, starting from the same
point and meeting later at some different point, so that the trajectories of
the two fluxons form together a closed loop. The non-trivial charge can move
as well. In this case
\begin{equation}
\gamma_{final} -\gamma_{initial}= 2\pi n\xi + 2\pi N,
\end{equation}
where $n$ is the winding number of the loop around the non-trivial charge
while $N$ represents the winding number of the loop around the bath
particles.

The result above contains the essence of our effect. We will give a number
of generalization later, but first let us make some comments.

The key element in our effect is the addition of the random bath of trivial
charges. When there are no trivial charges present, the effect is purely
local - monitoring the changes of the internal angle we can tell about the
presence of the non-trivial charge. The vector-potential generated by the
non-trivial charge, $\xi L \vec\nabla_R\theta$ is {\it ``gauge-invariant"}
and {\it observable}. As we add more and more trivial charges at random
positions and having random motion, the vector-potential generated by the
non-trivial charge becomes {\it unobservable}. The only gauge invariant
quantity becomes the ``loop integral", i.e. the change in the relative
internal angle over the closed loop. One can see a certain analogy between
the observability and non-observability in this case and the gauge
independence of the vector potential and its loop interal.

The two particle topological effect considered here is, up to a point,
analogous to an interference experiment with a single quantum particle such
as the Aharonov-Bohm experiment. The relative phase accumulated along two
trajectories in the quantum interference effect is hence analogous to the
relative internal angle in our case. There are however significant
differences. The first, and obvious difference, is that in the AB effect
there is a single particle, (whose wave-function is split in two
wave-packets), while in the classical analog we have two particles, each
following a well-defined classical trajectory. A more subtle difference is
the following. In quantum interference one is always sensitive to the
relative phase of two wave packets, since the measured quantity is the {\it
square} of the wave function. In the classical case, in contrast, we are
able to generalize our model to a situation where there are three particles,
with three internal angles, and where the only observable quantity involves
the internal angles of all three particles.

To illustrate this, we consider 3 fluxon-like particles which interact with
a single non-trivial source with a coupling strength $\xi$ located at the
origin, and with 3 {\em different} trivial random background charges denoted
by $A$, $B$, and $C$. The first fluxon sees particles of type $A$ as
positive charges and particles of type $B$ is negative charges. The second
fluxon sees type $B$ as a positive charge and and type $C$ as negative
charge and the third fluxon sees type $C$ as positive charges and type $A$
as negative charges.

The Hamiltonian of the system is then
\begin{eqnarray}
H_3={\frac{1}{2M_1}}[P_1 +L\nabla_{R_1}(\xi\theta_{1}+\sum_i(\theta_{1i}^A-
\theta_{1i}^B))]^2+  \nonumber \\
{\frac{1}{2M_2}}[P_2 +L\nabla_{R_2}(\xi\theta_{2}+ \sum_i(\theta_{2i}^B-
\theta_{2i}^C))]^2 +  \nonumber \\
{\frac{1}{2M_3}}[P_3 +L\nabla_{R_3}(\xi\theta_{3}+ \sum_i(\theta_{3i}^C-
\theta_{3i}^A)]^2 +  \nonumber \\
{\frac{1}{2m_A}}\sum_i[p_i^A+L\nabla_{r^A_i}(\theta_{i1}^A-\theta_{i3}^A)]^2+
\nonumber \\
{\frac{1}{m_B}}\sum_i[p_i^B+L\nabla_{r^B_i}(\theta_{i2}^B-\theta_{i1}^B)]^2+
\nonumber \\
{\frac{1}{2m_C}}\sum_i[P_i^C+L\nabla_{r^C_i}(\theta_{i3}^C-\theta_{i2}^C)]^2.
\end{eqnarray}

Let $\phi$ be the sum of the three internal angles $\phi=\varphi_1+%
\varphi_2+\varphi_3$. The change in the sum of the internal angles $%
\delta\phi$ which in the present model is "shared" by all 3 particles is
given by
\begin{eqnarray}
\delta\phi=\xi(\delta\theta_{1} +\delta\theta_{2} + \delta\theta_{3}) +
\sum_i(\delta\theta^A_{1i}-\delta\theta^A_{3i}) +  \nonumber \\
\sum_j(\delta\theta^B_{2j}-\delta\theta^B_{1j})+
\sum_k(\delta\theta^C_{3k}-\delta\theta^C_{1k}).
\end{eqnarray}
We note that the contribution of the last three sums over $i,j$ and $k$ is
random. However when the three fluxons start initially from the same point,
and end at the same final point, the random contribution exactly cancel
(modulo $2\pi$) and we are left with
\begin{equation}
\phi_{final}-\phi_{initial} = 2\pi \xi (n_1+n_2+n_3).
\end{equation}

Unlike the previous example here the change in $\phi$ yields the sum of the
winding numbers of each fluxon $n_1+n_2+n_3$.

The effects presented above are classical non-local analogs of quantum
vector-potential effects, such as the magnetic A-B effect and the
Aharonov-Casher effect. Along the same lines we now present an analog to the
scalar A-B effect. This is implemented by the interaction Hamiltonian
\begin{equation}
H_{int}=LV(x).
\end{equation}

In regions where the potential $V(x)$ is constant, this interaction doesn't
generate any {\it force}. Indeed, the force due to this interaction term is
equal to $F=-L{\frac{{dV}}{{dx}}}$ and it is zero where the potential is
constant. On the other hand, the internal angle $\varphi$ is affected: due
to the interaction it suffers an additional change of $V\Delta T$, where $%
\Delta T$ is the time spent in the potential $V$.

Again, in the absence of the randomizing charges background, the change of
the internal angle due to the potential is observable. However the
randomizing background makes the change in the internal angle unobservable.
An observable effect can be seen only in a ``closed loop" experiment similar
to that in the magnetic case.

In conclusion, we have described a classical non-local effect, analog to the
quantum Aharonov-Bohm and Aharonov-Casher effects. Although many classical
analogs to the AB and AC effects are known, they exhibit only the
topological character of the AB and AC effects but are local - by local
measurements one can see how the topological phase build up gradually. As
far as we know, this is the first classical non-local model which does not
involve general relativity and non-trivial space-time structures. In our
model, although one can measure at any time the internal angle of a
``fluxon", the measurement yields no information about a non-trivial charge.
Information can be obtained only in experiments in which a loop is closed.
The key ingredient which allows us to transform a local topological effect
into a non-local one is the addition of random but topologically trivial,
noise. A more detailed discussion of the issue of observability versus
unobservability in our model and its relations with cryptography are further
discussed in \cite{cryptography}.

Y. A., B.R. and A.S. acknowledge support from the Israel Science Foundation,
established by the Israel Academy of Sciences and Humanities. Y.A. and B.R.
are supported by the Israel MOD Research and Technology Unit. Y.A.
acknowledges support and hospitality of the Einstein center at the Weizmann
Institute of Science.


\begin{references}
\bibitem{point-Q}
Alternatively, the same effective Hamiltonian can be reached
by replacing the line of  charge with a point like charged source $Q$ and
the point-like magnetic fluxon with line of magnetic fluxons
(flux-line) that points in the fixed direction $\hat z$.
\bibitem{note1}  In the Hamiltonian (\ref{hl}) $\varphi$ is not an
independent variable, being fully determined by $\theta$. Employing a
Lagrangian analysis necessitates an addition of a kinetic term $L_z^2/2I$,
which removes the constraint of $\varphi$ to $\theta$ and leads to the well
defined Lagrangian
\begin{equation}
L= {\frac{1}{2}} m v^2 + {\frac{1}{2}} I(\dot\varphi - \vec v\cdot\vec
\nabla%
\theta)^2
\end{equation}
for every finite $I$. Our results are reproduced when the limit $%
I\rightarrow\infty$ is taken. We note that unlike the usual ${\vec v}\cdot
{%
\vec A}$ coupling, the Lagrangian we obtain has also a quadratic term of $%
v\cdot A$.
\bibitem{chargehamiltonian}  There should be, of course, a kinetic term for
the non-trivial charge as well. The term is absent here because, for
simplicity, we took this charge to be fixed at the origin i.e. to have
infinite mass.
\end{references}
\end{document}